\documentclass{camera}
\usepackage{graphicx}  

\begin{document}

%
\title{Optical Observations of GRB Afterglows}

%
\author{S. Covino}

%
\organization{INAF / Brera Astronomical Observatory, Via E. Bianchi 46, 23807, Merate (LC) - Italy}

\maketitle

\begin{abstract}
GRB afterglows are among the best examples of astrophysical sources requiring a true multiwavelength observational
approach. Radiation processes and main physical ingredients can only be disentangled with knowledge of their spectral and temporal properties through the largest possible band, i.e. from the X-ray down to radio. We now briefly review some of the 
most relevant observational findings recently obtained through optical observations.
\end{abstract}

%

\section{Introduction}

Optical (and NearInfraRed, NIR) observations are crucial for GRB afterglow studies. They contribute to multiwavelength study of afterglows making use of mature technologies guaranteeing high sensitivity.  Moreover,  already now and more and more in the near future, there will be specialized instruments able to compete with the capabilities of space-borne soft X-ray telescopes. On the other hand, optical observations are also among the best tools to observe the GRB host galaxies and the associated supernov\ae, thus allowing us to study the possible progenitors and in general the physical conditions driving the occurrence of GRBs. 

\section{GRB\,090423: Farther and Farther}

One of the most exciting recent results was definitely the measure of the redshift of GRB\,090423 (z $\sim 8.2$) \cite{Tanv09,Salv09} which of course confirms the reliability of the attractive perspective of using GRBs as probes of the early universe. It is the case to stress that this sounding result was secured by means of data obtained with the best available facilities as the ESO-VLT\footnote{http://www.eso.org} but also by a clever use of intermediate size telescopes with devoted instruments as the Amici prism equipping the Italian 3.5m TNG\footnote{http://www.tng.iac.es/news/2004/12/15/amici\_slitless/} or the simultaneous multiband imager GROND\footnote{http://www.mpe.mpg.de/~jcg/GROND/} equipping the 2.2m ESO-MPE telescope. The intrinsic brightness of most GRB afterglows at early times together with the delayed evolution due to cosmological time dilation makes intermediate size telescopes still competitive for these difficult but very fruitful researches.

\section{GRB\,09012: Early-Time Polarimetry}

Following the same line of reasoning, it is striking to note that the other main observational result obtained in 2009 was secured by observations carried out by the Liverpool 2.1m telescope equipped with the RINGO\footnote{http://telescope.livjm.ac.uk/Info/TelInst/Inst/RINGO/} polarimeter. It is an observationally well established result that GRB afterglows can be weakly polarized in the optical a few hours/days after the high-energy event \cite{Cov09}. This simple fact is already full of implications for the main GRB afterglow radiation process. The detection of early-time polarization, or the lack of it, can instead be a powerful diagnostic tool for understanding the nature of the outflow supposed to originate a GRB \cite{Lazz06}. RINGO and even more the upgraded RINGO2 instrument, is at present the best facility to study optical early-time polarization thanks in particular to the rapid pointing capabilities (about 2-3\,min) of the Liverpool telescope. A first attempt carried out for GRB\,060418 \cite{Mund07} only resulted in a rather shallow, although not meaningless, upper limit at about 8\%. For GRB\,090102, they \cite{Stee09} instead measured polarized emission at about 10\% level a few minutes after the high-energy event. A possible interpretation is that the optical radiation was produced in the reverse-shock region and therefore the rather high measured polarization could give a substantial support to the hypothesis that GRB outflows are driven by magnetic energy. However, the interpretation of this result in the framework of the internal/external shock model \cite{Pir99} is not straightforward \cite{Gend09}.

\section{Multiwavelength Afterglow Observations}

One of the most successful instruments of the \textit{Swift} satellite \cite{Gehr03} is the soft X-Ray Telescope (XRT) which has been able to secure uniform coverage in the 0.3-10\,keV band for most of the observed GRBs. Optical observations have usually been less continuous, allowing researchers to study only poorly sampled, and often with no spectral information, light curves.

Recently, however, the increased availability of robotic ground-based telescopes and better coordination among the observers allowed researchers to study well sampled and rich light curves at optical wavelengths too. It become possible, therefore, to study the shallow decay phase often observed at X-ray in the optical as for XRF\,080330 \cite{Guid09}. 

Dense multicolor optical observations allowed also to study with better reliability the problem of determining host galaxy dust induced extinction and compare them with columnar absorptions detected at the X-rays. The issue is far from being settled and there are convincing case with no extinction at all, XRF\,080330, and with well delineated SMC-like extinction curve as in GRB\,071010A \cite{Cov08}. A few cases with extinction resembling the one commonly observed in the Milky Way have also been singled out both by means of photometry and spectroscopy. One of the most recently published is GRB\,070802 with spectroscopical observations obtained with the ESO-VLT \cite{Elia09} and multicolor photometry with GROND \cite{Kruh08a}.

As a matter of fact, the GROND instrument is probably the best available instrument for studying GRB afterglow light curves providing wide band (from the $U$ to the $K$ band) and truly simultaneous observations. By means of GROND observations there is an increasing number of GRB afterglows with optical data of comparable if not of better quality than the simultaneous \textit{Swift}-XRT data. This allowed to study in some detail the early (starting from a few minutes after the high-energy event) afterglow or the occurrence of flares for several GRBs as XRF\,071031 \cite{Kruh08b}, GRB\,080129 \cite{Grei09} and GRB\,080710 \cite{Kruh09}. 

We finally mention in this section the impressive capabilities of the first second generation instrument available for the VLT: the X-shooter spectrograph\footnote{http://www.eso.org/sci/facilities/paranal/instruments/xshooter/} \cite{Cov06,Kap09} able to secure again truly simultaneous spectra from the $U$ to the $K$ band. The spectrum of GRB\,090313  \cite{deUg09} just allow us to infer the richness of information it will be possible to derive from X-shooter GRB afterglow spectra.

\section{Early- and Late-Time Observations}

The very early afterglow (up to a few minutes) is still kingdom of small size robotic telescopes while only exceptional events as GRB\,080319B \cite{Rac08} allow wide-field cameras as TORTORA or `Pi of the Sky"\footnote{http://grb.fuw.edu.pl/index.html} to provide useful data. Although most of the early time GRB afterglows are too faint for these telescopes, after several years of almost uninterrupted observations there are now remarkable databases of very early detections as those provided by ROTSE\footnote{http://www.rotse.net/} and TAROT\footnote{http://tarot.obs-hp.fr/} \cite{Ryk09,Klo09}. 

On the other hand, the late-time monitoring of GRB afterglows can only be afforded with the largest telescopes and in any case spending a substantial amount of telescope time. Contrary to some naive expectations late-time afterglows are sometime sources of unexpected behaviors difficult to reconcile with the leading internal/external shock scenario. A striking case is provided by the short GRB\,070707 \cite{Pira08} where ESO-VLT monitoring showed afterglow emission essentially switching off at about one day after the high-energy event. In addition, late-time monitoring of a few long GRBs \cite{Cen09} allowed to identify late-time jet-break transitions. This might imply that in at least a few cases jet-breaks were not detected simply because the light-curve monitoring did not last enough.

Deep observations of short GRB fields allowed to identify in most cases their host galaxies showing clear differences when compared to the host galaxy population of their longer duration siblings. Morphology analysis requires high spatial resolution and observations from space with the HST\footnote{http://www.stsci.edu/resources/} \cite{Fong10} are definitely the best tool although ground-based observations under very good conditions still play a role \cite{Dav09}.

Finally, the observations of GRB\,090426 \cite{Lev09,Ant09}, possibly a short GRB at redshift about 2.6, showed that short GRB formation could be feasible even at high redshifts.

\section{Conclusions}

Optical follow-up of GRB afterglows is still a key ingredient for afterglow studies. In the recent years there has been an increasing attitude by the observers to share their data producing papers with essentially all the available information. There are therefore several events with optical coverage comparable in quality to the \textit{Swift}-XRT soft X-ray coverage. In addition advanced facilities like GROND and RINGO, although equipping relatively small size telescopes, are providing exciting new data. In the near future it is likely that a new generation of intermediate size telescope will be designed, built and managed to allow a continuous and high quality coverage of GRB evolution, from a few seconds from the high-energy event down to the detection limits. The late-time evolution is also revealing to be a precious diagnostic and it will be followed by the largest facilities with better emphasis than in the past making a profitable use of the increasing number of 8m-class telescopes now operational.

%
\end{document}